\documentclass[epj]{svjour}
\usepackage{graphics}
\begin{document}
\title{Quantitative theory of channeling particle diffusion in transverse
energy in the presence of nuclear scarrering and direct evaluation
of dechanneling length }
\author{Victor V. Tikhomirov 
}                     
\offprints{Victor V. Tikhomirov}          
%
\institute{Institute for Nuclear Problems, \ Belarusian State
University, Minsk }
\date{Received: date / Revised version: date}
%
\abstract{ A refined equation for channe;ing particle diffusion in
transverse energy taking into consideration large-angle scattering
by nuclei is suggested. This equation is reduced to the
Sturm-Liouville problem allowing one to reveal both the origin and
the limitations of the dechanneling length notion. The values of
the latter is evaluated for both positively and negatively charged
particles of various energies. It is also demonstrated that the
dechanneling length notion is inapplicable for the nuclear
dechanneling process of positively charged particles, while the
effective electron dechanneling length of the latter vary more
than twice converging to a constant assymptotic value at the depth
exceeding the latter.
} 
\titlerunning{Channeling particle diffusion in transverse energy}
\maketitle
\section{Introduction}
\label{intro} Channeling effect in crystals delivers unique
possibilities of both high energy charged particle radiation and
control. Both electron and positron channeling makes it possible
to devise the new semi-mono\-chromatic sources of x- and
$\gamma$-radiation \cite{Bar,Bar2,Kor}. Proton and ion planar
channeling in bent crystals is a promising tool for both
extraction and collimation of the beams of the Large hadron
collider (LHC) and Future circular collider (FCC)
\cite{Sca,Bir,Zim}. Planar channeling of charmed and beauty
baryons in bent crystals also makes it possible to study the
effects of both CP- and CPT violation \cite{Bot}.

All the applications of channeling effect are limited by its
instability, induced by the dechanneling process. To describe the
latter, a concept of particle diffusion in the energy of its
transverse motion (transverse energy) was suggested soon after the
channeling discovery \cite{Lin,Gem,Bel}.

The dechanneling process is often characterized by the visual
notion of dechanneling length. Some information concerning the
latter can be extracted from Monte Carlo simulations. However,
since the average dechanneling distance of an individual particle
strongly depends on its initial transverse energy, any direct
method of dechanneling length evaluation through either the
channeling fraction or dechanneling distance averaging over the
incident particle angular distribution will give a result
depending on the latter. In addition, there is no ground to expect
that a channeling fraction, evaluated by any averaging method,
will exponentially depend on the particle penetration depth,
justifying an introduction of a dechanneling length independent of
the latter.

In fact, only the theory \cite{Bel}, describing the collective
properties of statistical particle behavior and consisting in
finding the lowest eigen number of the diffusion equation, can be
used for both strict introduction and evaluation of the
dechanneling length. Since the dechanneling lengths of positively
charged particles reach meters and tens of them at the LHC and FCC
energies, the direct method \cite{Bel} of their evaluation is more
superior than the Monte Carlo simulations which become time
consuming in the TeV particle energy region.

The theory \cite{Bel} is applicable only in the case of electron
dechanneling of nonrelativistic ions. Since the maximal angle
$\theta_{max} = m/M$ of scattering by electrons considerably
exceeds the critical channeling angle of non-relativistic ions,
the electron dechanneling of the latter can be correctly described
in neglect of both large-angle catastrophic scattering and the
scattering angle fourth power contribution to the mean square
variation of transverse energy. However, these assumptions loose
their applicability in many other cases.

First of all, since the channeling angle decreases with energy
becoming much less than $\theta_{max} = m/M$, the modification
\cite{Bir2,Bir} of the theory \cite{Bel} for the ultrarelativistic
case does not take into consideration adequately the particle
scattering by crystal atom electrons. Also nuclear scattering both
limits the fraction of channeling positively charged particles and
is essential for channeling of negatively charged ones,
considerably complicating the latter by the rechanneling process
\cite{Tik,Maz}.

Recently the experiments with bent crystals were first conducted
to separate the nuclear dechanneling process of both positively
\cite{Sca2,Bag} and negatively \cite{Sca3} charged particles. The
observed dechanneling fraction was fitted by the exponential decay
law being necessary to introduce a constant nuclear dechanneling
length. However, neither a justification of introduction nor a way
of evaluation of the latter were suggested in
\cite{Sca2,Bag,Sca3}.

To provide a correct evaluation of the dechanneling length in the
presence of nuclear scattering, an improved diffusion equation,
which takes into consideration both the scattering angle fourth
power contribution to the mean square variation of transverse
energy and large-angle catastrophic scattering, is introduced in
this paper. This equation is used to evaluate to dechanneling
length for the largest accelerator energies as well as to reveal
the peculiarities of the dechanneling process introduced by both
nuclear and electron scattering of both positively and negatively
charged particles.

\section{Refined equation for channeled particle diffusion in transverse energy}
\label{sec:1} \subsection{New features of the diffusion equation
in transverse phase space} 
\begin{figure}
\resizebox{0.48\textwidth}{!}{%
  \includegraphics{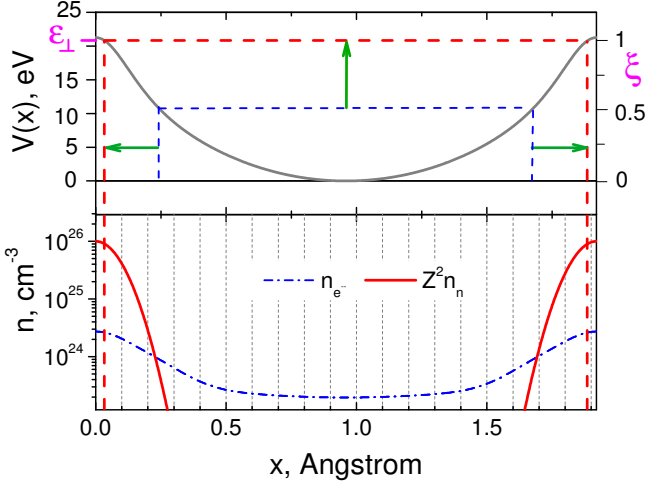}
} \caption{Planar potential and the undertaken expansion of the
considered region of positively charged particle motion, shown by
arrows (top). Averaged number density of electrons and the same,
multiplied by the atomic number squared, of nuclei (bottom). }
\label{fig:1}       
\end{figure}

According to the Lindhard averaged
potential concept \cite{Bar,Bir,Lin,Gem}, particle motion at small
angles with respect to crystal planes is described by the averaged
atomic potential $V(x)$ (potential energy, see Fig. \ref{fig:1})
which translation symmetry justifies the introduction of the
conserving energy of transverse motion \footnote{The system of
units $\hbar = c = 1$ is used.}
\begin{equation} \label{eq1} \varepsilon _ \bot =
\varepsilon v_x^2 / 2 + V(x) = p_x^2 / 2\varepsilon + V(x), \quad
\end{equation}
or transverse energy for short, in which $\varepsilon$ is total
particle energy, $p_x=\varepsilon v_x$ and $v_x (\varepsilon _
\bot ,x) = \sqrt {2\left[ {\varepsilon _ \bot - V(x)} \right] /
\varepsilon } $ are, respectively, its momentum and velocity
projections on the x axis normal to the crystal planes.
Conservative particle motion in the potential V(x) is always
disturbed by the incoherent scattering by both nuclei and
electrons  -- see Fig. \ref{fig:1}. At that, an instant incoherent
deflection by the angle $\theta_x$ in the point $x$ induces a
transverse energy change from (1) to
\begin{equation} \label{eq2} \varepsilon' _\bot =
\varepsilon [v_x(x) + \theta_x]^2 / 2 + V(x) = \varepsilon_\bot +
\varepsilon v_x(x) \theta_x + \varepsilon \theta^2_x/2.
\end{equation}
To describe the cumulative result of such changes, Fokker-Planck
approximation  \cite{Bar,Bir,Lif,Bai}, which, is able to treat
small transverse energy changes only \cite{Bel,Bir,Bai}, is widely
applied. Following the well established procedure \cite{Lif} and
using the notations of \cite{Bai}, one can introduce a
distribution function in the one dimensional transverse phase
space
\begin{equation}
\label{eq3} F(\varepsilon _ \bot ,x,z) = \varphi (\varepsilon _
\bot ,z)f_{\varepsilon _ \bot } (x)
\end{equation}
and the same over the transverse energy
\begin{equation}
\label{eq4} \varphi (\varepsilon _ \bot ,z) \equiv
\frac{1}{N}\frac{dN}{d\varepsilon _ \bot },
\end{equation}
where
\begin{equation}
\label{eq5} f_{\varepsilon _ \bot } (x) = \frac{2}{T\,v_x
(\varepsilon _ \bot ,x)}.
\end{equation}
is the coordinate space distribution function in which
\begin{equation}
\label{eq6} T(\varepsilon_ \bot ) = 2\int\limits_{x_l
(\varepsilon_ \bot )}^{x_r (\varepsilon_ \bot )} {\frac{dx}{v_x
(\varepsilon _ \bot, x)}} ,
\end{equation}
$x_l (\varepsilon_ \bot)$ and $x_r (\varepsilon_ \bot)$ are the
period and the right and left turning points of the channeling
motion at given $\varepsilon_ \bot$. The dependence of the phase
space distribution function (\ref{eq3}) on the depth $z$ of
particle penetration into the crystal along the channeling planes
is governed by the equation
\begin{equation}
\label{eq7} \frac{\partial F}{\partial z} = - \frac{\partial
}{\partial \varepsilon _ \bot }\left( {\frac{\Delta \varepsilon _
\bot }{\Delta z}F} \right) + \frac{1}{2}\frac{\partial
^2}{\partial \varepsilon _ \bot ^2 }\left( {\frac{(\Delta
\varepsilon _ \bot )^2}{\Delta z}F} \right) - wF,
\end{equation}
\begin{equation}
\label{eq8} \frac{\Delta \varepsilon _ \bot }{\Delta z} =
\frac{\Delta \varepsilon _ \bot (\varepsilon _ \bot ,x)}{\Delta z}
= \int (\varepsilon'_ \bot - \varepsilon _ \bot ) d \Sigma,
\end{equation}
\begin{equation}
\label{eq9} \frac{(\Delta \varepsilon _ \bot )^2}{\Delta z} =
\frac{(\Delta \varepsilon _ \bot )^2 \left( \varepsilon _ \bot ,x
\right)}{\Delta z} = \int {({\varepsilon '}_ \bot - \varepsilon _
\bot )^2d{\kern 1pt} \Sigma} ,
\end{equation}
which was deduced following \cite{Bai} and supplemented here for
the first time by the last term, which contains the probability
\begin{equation}
\label{eq10} w = w\left( \varepsilon _ \bot ,x \right) = \int' {d
\Sigma}
\end{equation}
of "catastrophic" scattering and describes the single scattering
process of particle immediate knocking out from the channeling
state. Another new feature will be the preservation of the forth
scattering angle power contribution to the integrand of Eq.
(\ref{eq9}), to introduce which a specific definition of the
integral regions of Eq. (\ref{eq8}) -(\ref{eq10}) is introduced
below.

To make the consideration more transparent, we will use the
simplified expression for the particle macroscopic scattering
cross section on both nuclei with local number density $n_n (x)$
and electrons with local number density $n_e (x)$:
\begin{equation}
\label{eq11} d\Sigma = \frac{4 \alpha^2 [Z^2 n_n (x) + n_e
(x)]}{\beta ^2p^2 (\theta^2 + \theta _1^2 )^2}d\theta_x d\theta_y,
\end{equation}
where $\theta_1$ is the angle, which takes the atomic nucleus
potential screening into consideration. Both $Z^2 n_n (x)$ and
$n_e (x)$ coordinate dependence is illustrated by Fig.
\ref{fig:1}. The integrand of the mean-squared transverse energy
variation rate (9)
\begin{equation}
\label{eq12} ({\varepsilon '}_ \bot - \varepsilon_\bot )^2 = 2
\varepsilon \left[\varepsilon_\bot - V(x) \right] \theta_x^2 +
\varepsilon^2 \theta_x^4/4 + .. ,
\end{equation}
in which the odd powers of the scattering angle $\theta_x$ are
omitted, contains the fourth power contribution of the same, which
has never been taken into consideration before
\cite{Bel,Bir2,Bir}.  The point is that, as can be directly seen
from Eqs. (\ref{eq11}) and (\ref{eq12}), this contribution
diverges at large $\theta_x$ and to treat it, finite integration
limits should be introduced. To define them, the transverse energy
(\ref{eq2}) of the scattered particle has been equated to the
height $V_{max}$ of the planar potential.The boundary scattering
angles which follows
\begin{equation}
\label{eq13} \theta _\pm (\varepsilon_\bot ,x) = - v_x
(\varepsilon_\bot ,x)\pm \sqrt {2\left( {V_{max} - V(x)} \right) /
\varepsilon}
\end{equation}
limit the integration region in Eq. (\ref{eq8}) and (\ref{eq9}) by
the angles $\theta_ - (\varepsilon_\bot ,x) \le \theta_x \le
\theta_+ (\varepsilon_\bot ,x)$, which leave the channeling
particles channeled, and fix the complimentary integration regions
$\theta_x
> \theta_ + (\varepsilon_ \bot ,x)$, $\theta_x < \theta_ -
(\varepsilon_ \bot ,x)$ in Eq. (\ref{eq10}). The integration over
all the angles $\theta_y = \sqrt{\theta^2 - \theta_x^2}$ of
scattering in the plane $yz$ is assumed everywhere in Eqs.
(\ref{eq8})-(\ref{eq10}).

Given the integration limits explicitly determined, the integrals
(\ref{eq8})-(\ref{eq10}) can be routinely taken with the result
\begin{equation}
\label{eq14}
\begin{array}{l}
\frac{\Delta \varepsilon_ \bot (\varepsilon_ \bot ,x)}{\Delta z} =
\frac{\pi \alpha^2}{\beta^3p}[Z^2n_n (x) + n_e (x)] \\
\times \left\{ \ln \left[ \frac{\theta_ + (x) + \sqrt {\theta_ +
^2 (x) + \theta_1^2 } }{\theta_ - (x) + \sqrt {\theta_ - ^2 (x) +
\theta_1^2 } } \right] \right. \\ \left. + \frac{\theta_ -
(x)}{\sqrt {\theta_ - ^2 (x) + \theta_1^2 } } - \frac{\theta_ +
(x)}{\sqrt {\theta_ + ^2 (x) + \theta_1^2 } } \right\}
\end{array}
\end{equation}
for the rate of transverse energy variation growth;
\begin{equation}
\label{eq15} \frac{(\Delta \varepsilon _\bot )^2(\varepsilon_\bot
,x)}{\Delta z} = a(\varepsilon_\bot,x) + b(\varepsilon_\bot,x),
\end{equation}
where
\begin{equation}
\label{eq16} a(\varepsilon_\bot ,x) = 4\left[ {\varepsilon_\bot -
V(x)} \right]\frac{\Delta \varepsilon_\bot }{\Delta z}
\end{equation}
and
\begin{equation}
\label{eq17}
\begin{array}{l}
 b(\varepsilon_\bot ,x) = \frac{\pi \alpha^2}{4}[Z^2n_n (x)
+ n_e (x)]\\
\times\left\{ {\theta_ + (x)\sqrt {\theta_+ ^2 (x) + \theta_1^2 }
}
 - \theta_ - (x)\sqrt {\theta_-^2 (x) +
\theta_1^2 } \right. \\
 \left.  + \frac{2 \theta_1^2 \theta_+
(x)}{\sqrt {\theta_+^2 (x) + \theta_1^2 } } - \frac{2\theta_1^2
\theta_- (x)}{\sqrt {\theta_-^2 (x) + \theta_1^2 } } \right. \\
\left. - 3\theta_1^2 \ln \left[ {\frac{\theta_ + (x) + \sqrt
{\theta_ + ^2 (x) + \theta_1^2 } }{\theta_ - (x) + \sqrt {\theta_
- ^2 (x) + \theta_1^2 } }} \right] \right\}
 \end{array}
\end{equation}
for the rate of squared transverse energy variation growth and

\begin{equation}
\label{eq18}
\begin{array}{l}
 w(\varepsilon_ \bot ,x) =
\frac{\pi \alpha^2}{\beta^2 p^2 \theta_1^2 }[Z^2 n_n(x) + n_e (x)]
\\ \times \left\{ 2 + \frac{\theta_-(x)}{\sqrt {\theta_-^2 (x) + \theta_1^2 }} - \frac{\theta_ + (x)}{\sqrt {\theta_-^2 (x) + \theta_1^2 }} \right\}
\end{array}
\end{equation}
for the catastrophic scattering probability.

\subsection{Reduction of the diffusion equation to the transverse
energy space} To reduce the diffusion equation (\ref{eq7}) in the
transverse phase space to that in the transverse energy space, the
averaging over the period of transverse motion
\begin{equation}
\label{eq19} \left\langle {\Phi (\varepsilon_ \bot ,x)}
\right\rangle = \int\limits_{x_l (\varepsilon_ \bot )}^{x_r
(\varepsilon_ \bot )} {\Phi (\varepsilon_ \bot ,x)f_{\varepsilon_
\bot } (x)dx}
\end{equation}
is used \cite{Bai} resulting in equation
\begin{equation}
\label{eq20}
\begin{array}{l}
\frac{\partial \varphi (\varepsilon_ \bot ,z)}{\partial z} = -
\frac{\partial }{\partial \varepsilon_ \bot }\left(
{A(\varepsilon_ \bot )\frac{\partial }{\partial \varepsilon_ \bot
}\frac{\varphi (\varepsilon_ \bot ,z)}{T(\varepsilon_ \bot )}}
\right) \\
 + \frac{\partial ^2}{\partial \varepsilon_ \bot ^2
}\left( {B(\varepsilon_ \bot )\frac{\varphi (\varepsilon_ \bot
,z)}{T(\varepsilon_ \bot )}} \right) - W(\varepsilon_ \bot
)\varphi (\varepsilon_ \bot ,z),
\end{array}
\end{equation}
containing the averaged coefficients
\begin{equation}
\label{eq21}
\begin{array}{l}
A(\varepsilon_ \bot ) = \left\langle {\frac{\Delta \varepsilon_
\bot (\varepsilon_ \bot ,x)}{\Delta {\kern 1pt} z}} \right\rangle
, \\ B(\varepsilon_ \bot ) = \left\langle {b(\varepsilon_ \bot
,x)} \right\rangle , \\ W(\varepsilon_ \bot ) = \left\langle
{w(\varepsilon_ \bot ,x)} \right\rangle.
\end{array}
\end{equation}
Eq. (\ref{eq20}) suggests, instead of transverse energy, to
introduce both the variable
\begin{equation}
\label{eq22} \xi'(\varepsilon_ \bot ) =
\int\limits_0^{\varepsilon_ \bot } {T(\varepsilon_ \bot
)d\varepsilon_ \bot },
\end{equation}
and corresponding distribution function
\begin{equation}
\label{eq23} u(\xi) = \frac{\varphi (\varepsilon_ \bot
)}{T(\varepsilon_ \bot )} = \frac{1}{N}\frac{dN}{T(\varepsilon_
\bot )d\varepsilon_ \bot } = \frac{1}{N}\frac{dN}{d \xi'}.
\end{equation}
The variable (\ref{eq22}) has a clear semiclassical meaning, being
equal to the multiplied by $2\pi$ quantum number of transverse
oscillatory motion in the state corresponding to the considered
transverse energy $\varepsilon_\bot$. Both Eqs. (\ref{eq22}),
(\ref{eq23}) and the standard transformations \cite{Mat} allows
one to extract further the Sturm-Liouville operator in Eq.
(\ref{eq20})
\begin{equation}
\label{eq24} r(\xi')\frac{\partial u(\xi',z)}{\partial z} =
\frac{\partial }{\partial \xi'}\left( {p'(\xi')\frac{\partial
u(\xi',z)}{\partial \xi'}} \right) - q(\xi')u(\xi',z)\;,
\end{equation}
with the coefficients
\begin{equation}
\label{eq25} p'(\xi') = \left[ B \left( \varepsilon_ \bot (\xi')
\right) + A\left( \varepsilon_\bot (\xi') \right) \right] T \left(
\varepsilon_ \bot (\xi') \right)r(\xi'),
\end{equation}
\begin{equation}
\label{eq26} q(\xi' ) = \left[ {W\left( \varepsilon_ \bot (\xi')
\right) - {B}''\left( \varepsilon_ \bot (\xi') \right)} \right]T^{
- 1}\left( {\varepsilon_ \bot (\xi')} \right)r(\xi')
\end{equation}
and
\begin{equation}
\label{eq27}
 \begin{array}{l}
  r(\xi') = \exp \int\limits_0^{\varepsilon_\bot (\xi')}
{\frac{{B}'(\varepsilon_ \bot )d\varepsilon_ \bot }{A(\varepsilon_
\bot ) + B(\varepsilon_ \bot )}\,} , \\ \varepsilon_ \bot (\xi') =
\int\limits_0^{\xi'} {\frac{d\xi}{T(\varepsilon_\bot (\xi) )}\;},
 \end{array}
\end{equation}
which can be readily calculated using Eqs.
(\ref{eq14})-(\ref{eq19}) and (\ref{eq21}). It should be
emphasised that Eq. (\ref{eq24}) is more general than that used in
\cite{Bel,Bir2}, which follows from Eq. ((\ref{eq24}) at $q=0$ and
$r=const$.

\subsection{Diffusion equation boundary conditions}
Introducing the boundary conditions for Eq. (\ref{eq24}), we
immediately adopt the one of $\partial u(0)/\partial\xi' = 0$,
reflecting the impossibility of both transverse energy and
variable (\ref{eq22}) to decrease below zero. However, another
condition of the distribution function (\ref{eq23}) nullification
at some $\varepsilon_{\bot max}$ or $\xi'_{max}$, essential for
the present approach \cite{Bel,Bir2}, needs some comments. Indeed,
at first glance, the region of large $\varepsilon_\bot$ is surely
well populated by the intensively scattering dechanneling
particles. However, the diffusion equation is applicable in the
limit of small changes of the considered quantity, transverse
energy in our case. That is why one should adopt that at some
$\varepsilon'_\bot$ or $\xi'_{max}$, when the variation
\begin{equation} \label{eq28}
\begin{array}{l}
\delta \varepsilon_ \bot (\varepsilon') = \left( \left\langle
\frac{(\Delta \varepsilon_ \bot )^2(\varepsilon'_ \bot,x)}{\Delta
z} \right\rangle T(\varepsilon') \right. \\ \left. - \left\langle
\frac{\Delta \varepsilon_ \bot (\varepsilon'_ \bot,x)}{\Delta z}
\right\rangle ^2T^2(\varepsilon'_ \bot) \right)^{1 / 2}
\end{array}
\end{equation}
of the former over the channeling period  reaches the interval
$V_{max} - \varepsilon'_\bot$, separating $\varepsilon'_\bot$ from
the potential maximum, equation (\ref{eq24}) ceases to describe
any particle, justifying thus the second boundary condition
$u(\xi'_{max}) =0$, where $\xi'_{max} = \xi'(\varepsilon_{\bot
max})$ and $\delta\varepsilon_\bot(\varepsilon_{\bot max}) =
V_{max} - \varepsilon_{\bot max}$. To estimate the uncertainly of
this definition of $\varepsilon_{\bot max}$, we took the ratios
$$\delta\varepsilon_\bot(\varepsilon_{\bot max}) / \left[V_{max} -
\varepsilon_{\bot max}\right] = 0.5,~1,~2,$$ to demonstrate in
Table \ref{tab:1} that an uncertainty of the dechanneling length
definition is marginal.

Given the boundary value $\xi'_{max}$, one can redefine Eq.
(\ref{eq24}) to formulate Sturm-Liouville problem on the interval
$[0,1]$ of the normalized variable ${\xi } = \xi' / \xi'_{max}$ by
omitting the prime in Eqs. (\ref{eq23})-(\ref{eq27}) and putting
$p'(\xi) = p(\xi) \xi^{\prime 2}_{max}$:
\begin{equation} \label{eq29}
r(\xi)\frac{\partial u(\xi,z)}{\partial z} = \frac{\partial
}{\partial \xi}\left( {p(\xi)\frac{\partial u(\xi,z)}{\partial
\xi}} \right) - q(\xi)u(\xi,z)\;,
\end{equation}
\begin{equation}
\label{eq30}
 - \frac{\partial }{\partial \xi}\left[ {p(\xi )\frac{\partial
}{\partial \xi }u_n (\xi )} \right] + q(\xi )u_n (\xi ) = \lambda
_n r(\xi )u_n (\xi),
\end{equation}
\begin{equation}
\label{eq31}
\partial u_n (0) / \partial \xi = 0,\quad u_n (1) = 0, \quad
n=1,2,..,
\end{equation}
where we follow the sign convention of \cite{Pru}. The dependence
of the Eqs. (\ref{eq29}), (\ref{eq30}) coefficients on the
normalized parameter $\xi$ is illustrated by Fig. 2 in the case of
400 GeV protons and (110) Si planes, also used as an example in
Figs. \ref{fig:3}-\ref{fig:7} below.

\begin{figure}
\resizebox{0.4\textwidth}{!}{%
  \includegraphics{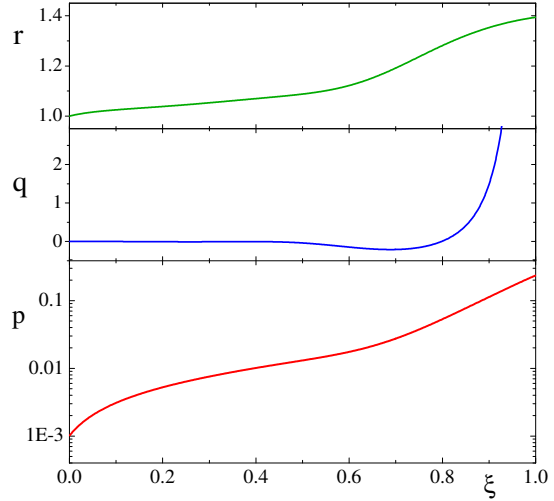}
} \caption{The Eqs. (29) and (30) coefficient dependence on the
parameter $\xi$ for 400 GeV protons channeled by the (110) Si
planes. }
\label{fig:2}       
\end{figure}

\section{Diffusion equation solution and its analysis}
\label{sec:2}
\subsection{Dechanneling length at different energies}
\begin{figure}
\resizebox{0.4\textwidth}{!}{%
  \includegraphics{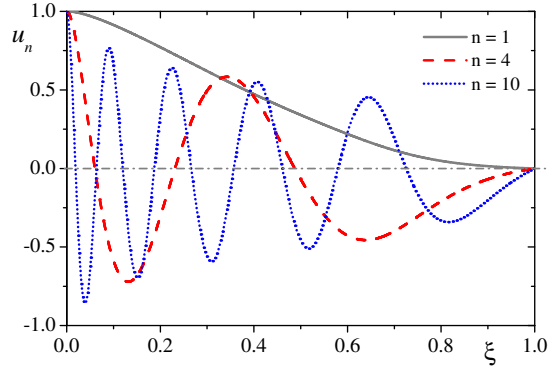}
} \caption{The 1-t, 4-th and 10-th eigen states for the same. }
\label{fig:3}       
\end{figure}
\begin{figure}
\resizebox{0.4\textwidth}{!}{%
  \includegraphics{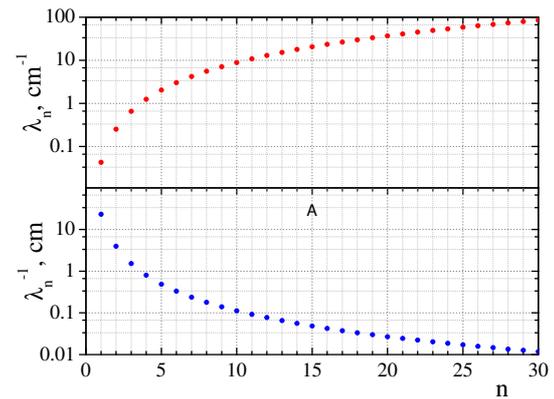}
} \caption{The eigen values of Eq. (\ref{eq30}) (top) and
corresponding decay lengths (bottom).}
\label{fig:4}       
\end{figure}
A numerical solution of Sturm-Liouville problem (\ref{eq30}),
(\ref{eq31}), like that of \cite{Pru}, allows one to find any
number of its eigen states $u_n(\xi)$ and eigen values
$\lambda_n$, $n=1, 2,..$, some of the lowes of which are plotted,
respectively, in Figs. \ref{fig:3} and \ref{fig:4}. The eigen
values $\lambda_n$ are positive and monotonically increase with
their numbers $n$, equal to the number of corresponding eigen
state $u_n(\xi)$ nodes minus one. The completeness property of the
system of eigen states of the problem (\ref{eq30}), (\ref{eq31})
allows one to represent any solution of Eq. (\ref{eq29}) in the
form of expansion
\begin{equation}
\label{eq32} u(\xi,z) = \sum\limits_{n = 1}^\infty {c_n \exp
\left( { - \lambda _n z} \right)u_n (\xi)},
\end{equation}
the coefficients $c_n$ of which are determined by the distribution
$u(\xi,0)$ at the crystal entrance $z=0$ and are evaluated below.
The solution (\ref{eq32}) allows one to determin the channeling
probability dependence on crystal depth
\begin{equation}
\begin{array}{l}
\label{eq33} P_{ch}(z) \equiv \frac{N_{ch} (z)}{N_0 } =
\int\limits_0^1 {u(\xi,z)d\xi} = \frac{1}{N_0 }\sum\limits_{n =
1}^\infty N_{n0} \exp \left( { - \lambda_n z} \right) \\
= \sum\limits_{n = 1}^\infty {c_n \bar{u}_n \exp \left( { -
\lambda_n z} \right) \;\buildrel {z\lambda_1 >1} \over
\longrightarrow } \;c_1 \bar{u}_1 \exp \left( { - \lambda_1 z}
\right),
\end{array}
\end{equation}
where
\begin{equation}
\label{eq34} \bar{u}_n = \int\limits_0^1 {u_n (\xi)d\xi}.
\end{equation}
An exponential decay of the eigen states is governed by their
eigen values, the smallest first of which alone determines the
asymptotic exponential behavior of the general solution
(\ref{eq32}), which gave rise to the introduction of the
dechanneling length $l_{dech} = 1/\lambda_1$ in \cite{Bel}.
Uncertainties of the latter, connected with both the qualitative
nature of the introduction of the boundary $\varepsilon_{\bot
max}$ of the diffusion approximation applicability region and the
planar potential model, are displaced in table 1 for 400 GeV
protons and (110) Si planes. As one can see, both the uncertainty
of the boundary energy $\varepsilon_{\bot max}$ definition and the
transition between the potential models \cite{Doy} and \cite{Tob}
change the $l_{dech}$ value by less then one percent. Considerably
smaller dechanneling length, obtained with the Moliere potential
\cite{Gem}, reflects the limited applicability of the latter to
the channeling phenomenon. In any case, the precision of
$l_{dech}$ determination by the diffusion equation method is not
worse than the uncertainty related with the potential model
choice.

\begin{table}
\caption{Dechanneling length and precision of its evaluation}
\label{tab:1}       
\begin{tabular}{lllll}
\hline\noalign{\smallskip} Potential & $\frac{\delta
\varepsilon_\bot (\varepsilon_{ \bot \max } )}{V_{\max } -
\varepsilon_{ \bot \max } }$& $l_{dech},~cm$ &
$\Delta l_{dech}$,{\%}&  \\
\noalign{\smallskip}\hline\noalign{\smallskip} Tob [14]& 1&
\textbf{23.1}& \textbf{0} \\
Tob [14]& 0.5& 22.9& -0.81 \\
Tob [14]& 2& 23.2& +0.37 \\
DT [15]& 1& 23.3& +0.61 \\
Mol [7]& 1& 21.4& -7.255 \\
\noalign{\smallskip}\hline
\end{tabular}
\end{table}
\begin{table}
\caption{Dechanneling lengths for protons and electrons of
different energies}
\label{tab:2}       
\begin{tabular}{llllll}
\hline\noalign{\smallskip} $e^-/p$ & $\varepsilon ,GeV$&
$l_{dech}$, cm & $\lambda_2 / \lambda_1 $& $\Delta l_{dech}$,
{\%} & $N_{ch0}/N_0$  \\
\noalign{\smallskip}\hline\noalign{\smallskip}
p &  400 &  23,1 & 6.0 &  0.61 &  0,895  \\
p & 6500 & 303.6 & 5.7  & 0.34 & 0.895  \\
p & $10^5$ & 3936.0 & 5.6 & 0.18 & 0.895 \\
$e^-$ & 1 & 6.0 $10^{-4}$ & 7.8 & 130.0 & 0.33 \\
$e^-$ & 10 & 50.0 $10^{-4}$ & 6.9 & 78.0 & 0.39  \\
$e^-$ & 100 & 0.044 & 6.4 & 46.0 & 0.44  \\
$e^-$ & 1000 & 0.38 & 6.1 & 28.5 & 0.49  \\
\noalign{\smallskip}\hline
\end{tabular}
\end{table}
Table \ref{tab:2} displays the $l_{dech}$ values for both
positively and negatively charged ultra-relativistic particles
demonstrating the large value of the ratio $\lambda_2/ \lambda_1
\sim 6 \div8 $, assuring the strong dominance of the first eigen
state starting from  $z \sim l_{dech}$ (see below).

\subsection{Nuclear dechanneling probability dependence on particle penetration depth }
\begin{figure}
\resizebox{0.4\textwidth}{!}{%
  \includegraphics{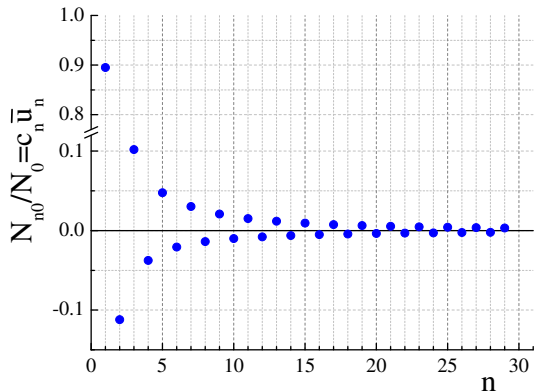}
} \caption{The eigen state amplitudes entering Eq. (\ref{eq33}).}
\label{fig:5}       
\end{figure}

However the evolution of the probability (\ref{eq33}) at the
smaller depths $z \leq l_{dech}$ is governed by many states,
making its behavior more complex. To study an interference of
different eigen states, one should know their amplitudes $c_n
\bar{u}_n$ (see Eq. (\ref{eq33})). The simplest formula can be
obtained for them in the limit of zero particle incidence, when $|
d\varepsilon_\perp | = | dV(x)/dx | dx$ and the distribution
function at the crystal entrance reduces to
\begin{equation}
\label{eq35} u(\xi,0) = \frac{1}{N}\frac{dN}{d\xi} =
\frac{\xi'_{max}}{d_{pl} T\left( {\varepsilon_ \bot (\xi)} \right)
\left| {dV\left( {x\left( {\varepsilon_ \bot (\xi)} \right)}
\right) / dx} \right|},
\end{equation}
allowing one to evaluate both the coefficients
\begin{equation}
\begin{array}{l}
\label{eq36} c_n = \int\limits_0^1 {u(\xi,0)u_n (\xi)r(\xi)d\xi}
\left( {\int\limits_0^1 {u_n^2 (\xi)r(\xi )d\xi} } \right)^{ -
1}\\= \int\limits_0^d u_n
\left(\xi(V(x))\right)r(\xi\left(V(x))\right) \frac{dx}{d} \left(
{\int\limits_0^1 {u_n^2 (\xi)r(\xi )d\xi} } \right)^{ - 1}
\end{array}
\end{equation}
of the expansion (\ref{eq32}) and the amplitudes $N_{n0}/N_0 = c_n
\bar{u}_n$ of the eigen states, which enter Eq. (\ref{eq33}) and
are plotted in Fig. \ref{fig:5}.

Eqs. (\ref{eq33}) and (\ref{eq36}) allow one to illustrate the
main features of the dechanneling process of positively charged
particles in the most practically important and complex region $l
< l_{dech}$. Since only the particles with sufficiently high
transverse energies reach the regions of atomic nuclei
localization (see Fig. \ref{fig:1}), the fast nuclear and slow
electron dechanneling processes, which dominate, respectively, at
sufficiently high and low transverse energies, can be
qualitatively distinguished. The question arises, however, is it
possible to strictly introduce and measure both the nuclear and
electron dechanneling lengths, as was done in \cite{Sca2,Bag}? It
is the knowledge of both the eigen values and the amplitudes
(\ref{eq36}) of eigen states, entering the solution (\ref{eq32}),
which make it possible to treat this question thoroughly.

\begin{figure}
\resizebox{0.4\textwidth}{!}{%
  \includegraphics{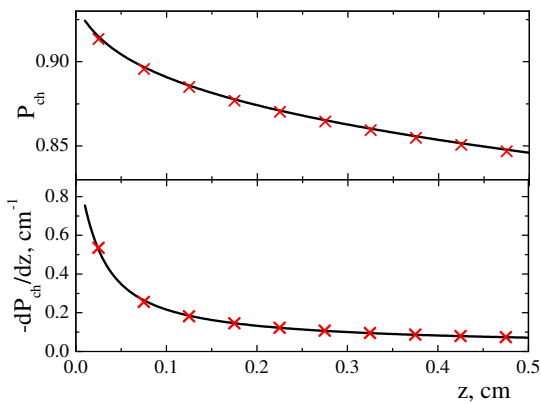}
} \caption{Crystal depth dependence of the channeling probability
(\ref{eq33}) (solid line, top) and dechnneling rate (solid line,
bottom) and their fits by the function (\ref{eq37}) and its
derivative, taken with the negative sign, respectively (crosses).}
\label{fig:6}       
\end{figure}

Fig.  \ref{fig:4} shows that the number of the eigen values, the
inverse values of which correspond to the typical nuclear
dechanneling region (about $1 mm$ at $400 GeV$), exceeds one
considerably. Corresponding eigen states also have the amplitudes,
comparable in value. The cumulative contribution of these multiple
and close fast decaying states is approximated by the integral of
the exponent $\exp(-\lambda (\varepsilon_\perp)z)$ product by a
slowly varying function of transverse energy well fitted by the
power-type
\begin{equation}
\label{eq37} P_{ch}(z) = \sum\limits_{n = 1}^\infty {c_n \bar{u}_n
\exp \left( { - \lambda_n z} \right) \;\buildrel {z \ll\lambda_1}
\over \longrightarrow }0.954 - 0.137\sqrt[3]{z},
\end{equation}
instead of exponential-type function of the crystal depth $z$, as
Fig. \ref{fig:6} demonstrates. Thus, in place of the dechanneling
length extraction, the experiments on nuclear dechanneling should
accept the power-type fiting of the channeling fraction dependence
on depth.
\begin{figure}
\resizebox{0.4\textwidth}{!}{%
  \includegraphics{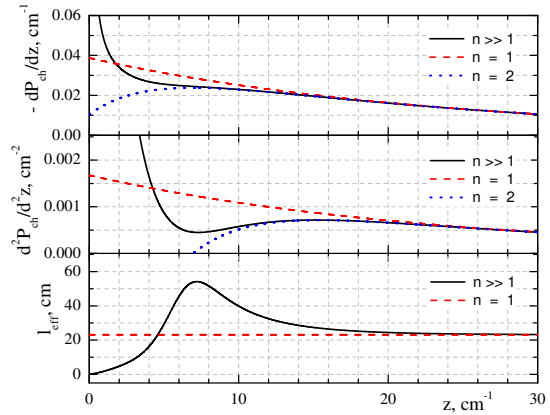}
} \caption{Dechanneling rate (\ref{eq38}), its derivative
(\ref{eq39}) and effective decahnneling length (\ref{eq40}) versus
crystal depth.}
\label{fig:7}       
\end{figure}

\subsection{Peculiarities of the electron dechanneling of positively charged particles}
On the opposite, a few lowest eigen values, describing the
electron dechanneling process, differ severalfold -- see Fig.
\ref{fig:4}. In addition, the smallest first of them possesses an
especially large amplitude $N_{10}/N_0 = c_1 \bar{u}_1$ -- see
Fig. \ref{fig:5}. By this reason the applicability of the electron
dechanneling length notion is assured starting just from $z \simeq
l_{dech} = 1/ \lambda_1$. However the real experiments on beam
steering, collimation and electromagnetic radiation generation are
conducted at $z < l_{dech}$ and even at $z \ll l_{dech}$, where
several lowest eigen state contribute considerably to both Eq.
(\ref{eq32}) and (\ref{eq33}), violating their asymptotic
exponential behavior. Indeed, let us consider both the
dechanneling rate
\begin{equation}
\label{eq38} -P'_{ch}(z) = \sum\limits_{n = 1}^\infty c_n
\bar{u}_n \lambda_n \exp \left( { - \lambda_n z} \right),
\end{equation}
which can be measured experimentally, and its derivative
\begin{equation}
\label{eq39} P''_{ch}(z) = \sum\limits_{n = 1}^\infty c_n
\bar{u}_n \lambda_n^2 \exp \left( { - \lambda_n z} \right)
\end{equation}
(see Fig. \ref{fig:7}). The point is that the relatively small
population coefficients $|c_n| \ll c_1$ with $n > 1$ are
multiplied by the large eigen values $\lambda_n \gg \lambda_1$ and
their squares in Eqs. (\ref{eq38}) and (\ref{eq39}), respectively,
making their "weights" comparable with that of the first eigen
state. Having, in addition to its high "weight", the negative sign
of the amplitude, the second eigen state directly distorts the
steady decrease of the leading contribution of the latter. To
demonstrate how the states with $n > 1$ modify the exponential
channeling decay law at $z < l_{dech}$, let us introduce the
effective dechanneling length
\begin{equation}
\label{eq40} l_{dech}^{eff}(z) = - \frac{P'_{ch}(z)}{P''_{ch}(z)}.
\end{equation}
The latter is equal to the constant quantity
$l_{dech}=1/\lambda_1$, either when only the first terms in the
sums (\ref{eq38}) and (\ref{eq39}) are preserved or if the region
$z \gg l_{dech}$ is considered. However at $z < l_{dech}$ the $n
> 1$ terms induce the considerable dependence of the effective
dechanneling length (\ref{eq40}) on depth, resulting in its
doubling at $z = 6\div8 cm$ -- see Fig. \ref{fig:7}. That is why,
instead of fitting the data by the single exponent containing a
constant electron dechanneling length, more complex behavior of
the dechanneling process of positively charged particles should be
considered to establish the the crystal thickness, optimal for the
applications. One of the most fundamental one of the channeling in
thick crystals is the measurement of both magnetic and electric
dipole momenta of short living particles \cite{Bot}.

\subsection{Diffusion equation application to the
dechanneling of negatively charged particles} Since the particle
scattering by nuclei is thoroughly taken into consideration by
Eqs. (\ref{eq29}) and (\ref{eq30}), the solution (\ref{eq32}) can
be also applied to the case of negatively charged particles.
Oppositely to the case of positively charged ones, all negatively
charged particles experience strong nuclear scattering, inducing
large fluctuations of transverse energy at any value of the
latter. As a result, more than 50\% of the particles (see table
\ref{tab:2}), having transverse energies $\varepsilon_{\bot max} <
\varepsilon_\bot \leq V_{max}$, experience average transverse
energy variations (\ref{eq28}) exceeding the depth $V_{max} -
\varepsilon_{\bot max}$ of their transverse energy level
occurrence. These variations induce either immediate dechanneling
or large changes of both the transverse motion period and phase,
making, in fact, inapplicable the whole notion of channeling,
understood as a quasiperiodic transverse motion. The rest of the
particles with $\varepsilon_\bot < \varepsilon_{\bot max}$, which
more likely can be considered as channeled, also experience large
transverse energy fluctuations resulting in the uncertainly of
dechanneling length which reaches several tens of percent -- see
table \ref{tab:2}.

Contrary to the case of positively charged particles, nuclear
scattering of negatively charged ones, which occurs near the
potential energy minimum, can immediately make their transverse
energy considerably less than the height of the potential barrier,
giving rise to the intensive reachanneling process which Eqs.
(\ref{eq29}), (\ref{eq30}) are unable to describe. Thus, in total,
the diffusion equation approach provides merely qualitative
information on the dechanneling process of negatively charged
particles. Taking into consideration that the dechanneling length
of negatively charged particles is much less than that of
positively charged ones (see table \ref{tab:2}), one should
acknowledge the Monte Carlo simulations to be a quite adequate
approach for negatively charged particle dynamics study. In
particular, this method correctly reproduces the nearly
exponential decay of the channeling population in a bent crystal
observed in \cite{Maz}, to describe  which the diffusion equation
should be additionally refined.

Monte Carlo simulations also certainly take into consideration all
possible features of positively charged particle motion. In
particular, they correctly describe both the rechanneling process
and the large transverse energy fluctuations at the upper
under-barrier region $\varepsilon_\bot \sim \varepsilon_{\bot
max}$, in which the diffusion equation looses its applicability.
Monte Carlo approach have also demonstrated its efficiency in
simulation of the new effects of channeling probability increase
by the crystal cut \cite{Tik2}, positive miscut influence of
collimation \cite{Tik3} and other possibilities of the latter
\cite{Syt}, multiple new radiation features in both crystal
undulators \cite{Bar2,Bag2,Tik4} and bent crystals \cite{gui,ban},
the effects of planar channeling and quasichanneling oscillations
in the deflection angle distribution of particles passed through a
bent crystal \cite{Syt2} and many effects in the field of atomic
strings \cite{tik5,Sca4,Sca5,ban3,gui2,ban2,tik6}. However the
main purpose of this paper was to demonstrate that despite all the
achievements of Monte Carlo method, the diffusion equation
approach, refined in present paper, can supplement the simulation
results with the enlightening treatment of the collective
statistical behavior of channeling particles, which can not be
described by the sum of two exponents decaying with nuclear and
electron dechanneling lengths.

Both electron and nuclear dechanneling lengths can be measured
using bent crystals \cite{Sca2,Bag,Sca3}. Crystal bending will
certainly change the dechanneling length values from table 2,
evaluated for a straight crystal.  Though both experiment and
simulations demonstrate reasonable applicability of thee
dechanneling length of electrons in bent crystals \cite{Maz}, to
predict its value diffusion equation (\ref{eq7}) should be further
expanded to take into consideration the strong rechanneling
effect. On the opposite, a marginal role of the rechanneling
process of positively charged particles allows one to assume that
the new conclusions concerning the dechanneling rate dependence on
crystal depth remain true also at the presence of crystal bending
and should be used to reinterpret the results of the experiments
of the \cite{Sca2,Bag,Sca3} type.

\section{Conclusions}
In this paper the equation of channeling particle diffusion in
transverse energy was supplemented with the effect of nuclear
scattering and applied for a direct evaluation of the huge
dechanneling length values at the multi-TeV particle energies. The
diffusion equation approach has also proved to be really
indispensable in revealing the general features of the collective
statistical behavior of channeling particles such as the
power-type channeling probability dependence on the particle
penetration depths in the nuclear dechanneling region and a strong
dependence of the effective electron dechanneling length on the
particle penetration depth in the interval between the nuclear
dechanneling region and approximately one electron dechanneling
length. These fundamental predictions, which strongly contradict
the straightforward application of both nuclear and electron
dechanneling lengths, should be used for an unbiased
interpretation of the simulation predictions and represent
themselves a challenge for the future experiments.

%
%

\end{document}